\documentclass{article}
\usepackage[T1]{fontenc} % add special characters (e.g., umlaute)
\usepackage[utf8]{inputenc} % set utf-8 as default input encoding
\usepackage{ismir,amsmath,cite,url}
\usepackage{graphicx}
\usepackage{color}
\usepackage[lined,boxed]{algorithm2e}
\usepackage{amsfonts}
\usepackage{booktabs}
\usepackage{multirow}

\usepackage{lineno}
\RestyleAlgo{ruled}

\newcommand{\RR}{\mathbb{R}}

\newcommand{\LL}{\mathcal{L}}

\def\x{{\mathbf x}}

\def\a{{\mathbf a}}

\newcommand{\floor}[1]{\lfloor#1\rfloor}
\newcommand{\ceil}[1]{\lceil#1\rceil}

\title{Singing Voice Synthesis Using Differentiable LPC and Glottal-Flow-Inspired Wavetables}

\multauthor
{Chin-Yun Yu \hspace{1cm} Gy{\"o}rgy Fazekas} 
{
Centre for Digital Music, Queen Mary University of London, UK \\
{\tt\small chin-yun.yu@qmul.ac.uk, george.fazekas@qmul.ac.uk}
}

\def\authorname{C.-Y. Yu and G. Fazekas}

\usepackage[bookmarks=false,pdfauthor={\authorname},pdfsubject={\papersubject},hidelinks]{hyperref}
\begin{document}

\maketitle
\begin{abstract}
% The abstract should be placed at the top left column and should contain about 150-200 words.
This paper introduces GlOttal-flow LPC Filter (GOLF), a novel method for singing voice synthesis (SVS) that exploits the physical characteristics of the human voice using differentiable digital signal processing.
GOLF employs a glottal model as the harmonic source and IIR filters to simulate the vocal tract, resulting in an interpretable and efficient approach. 
We show it is competitive with state-of-the-art singing voice vocoders, requiring fewer synthesis parameters and less memory to train, and runs an order of magnitude faster for inference. 
Additionally, we demonstrate that GOLF can model the phase components of the human voice, which has immense potential for rendering and analysing singing voice in a differentiable manner.
Our results highlight the effectiveness of incorporating the physical properties of the human voice mechanism into SVS and underscore the advantages of signal-processing-based approaches, which offer greater interpretability and efficiency in synthesis.
\end{abstract}
\section{Introduction}\label{sec:introduction}
Singing voice synthesis (SVS) has attracted substantial interest as a research topic over the last decades, and a variety of techniques have been developed.
Early successful SVS systems were usually based on sample concatenation~\cite{macon_singing_1997, bonada_singing_nodate, bonada_synthesis_2007, bonada_expressive_2016}, while parametric systems have become much more prevalent.
The actual synthesis process in parametric systems is carried out by a \emph{vocoder} controlled by synthesis parameters generated from a separate acoustic model given some musical context factors (i.e.\ note number, duration, phoneme, etc.).
Early systems of this kind use a linear source-filter model as vocoder~\cite{saino_hmm-based_2006, hono_recent_2018}. 
Deep Neural Networks (DNNs) have subsequently become the dominant approach for state-of-the-art vocoders~\cite{shen_natural_2017, valin_lpcnet_2019, yoneyama_unified_2022, prenger_waveglow:_2018, liu_diffsinger_2021, cho_survey_2021, takahashi_hierarchical_2022}.
However, mel-spectrograms are often chosen as input features to these models, which are less interpretable than traditional vocoder parameters (e.g.\ f0, aperiodicity ratios).
Also, a significant amount of data is needed to cover various vocal expressions to achieve generalisation.

In contrast, Differentiable Digital Signal Processing (DDSP) models~\cite{engel_ddsp_2020, hayes_neural_nodate, shan_differentiable_2021} incorporate existing signal processing operations into neural networks as an inductive bias, making them more interpretable and generalisable.
DDSP additive synthesis has been proposed for SVS by Alonso et al.~\cite{alonso_latent_2021}. Wu et al.~\cite{wu1_ddsp-based_2022} improved this further by using subtractive synthesis and sawtooth as the harmonic source.
Nercessian et al.~\cite{nercessian_differentiable_2022} proposed a differentiable version of the WORLD vocoder~\cite{morise_world_2016} for doing end-to-end singing voice conversion.
Yoshimura et al.~\cite{yoshimura_embedding_2022} used Taylor expansion to approximate the mel-log spectrum approximation filter's (MLSA) exponential function and embedded it into an SVS system.
However, most of their architectures only assume the target signal is a monophonic instrument, which can potentially lead to solutions that do not reflect some properties of voice.
In their design, the harmonic sources are fixed to a specific shape (e.g. sawtooth, pulse train), and the filters are symmetric in the time domain, except Yoshimura et al.~\cite{yoshimura_embedding_2022} which use a minimum-phase MLSA filter.
Incorporating constraints specific to the human voice on the harmonic source and the filters could lead to a more interpretable and compact SVS vocoder.

In this work, we propose GlOttal-flow LPC Filter (GOLF), an SVS module informed by the physical properties of the human voice.
We build upon the Harmonic-plus-Noise architecture of DDSP~\cite{engel_ddsp_2020} and the subtractive synthesis of SawSing~\cite{wu1_ddsp-based_2022}, but replace the harmonic source with a glottal model and use IIR filters.
We developed a differentiable IIR implementation in PyTorch~\cite{li_pytorch_2020} for training efficiency.
We then used this module as a neural vocoder and compared its performance with other DDSP-based vocoders.
Specifically, a simple and lightweight NN encoder converts the mel-spectrogram into synthesis parameters, and the synthesiser decodes the signal from it. 
We paired different synthesisers with the same encoder and trained them jointly.

Our contributions are twofold.
First, GOLF has significantly fewer synthesis parameters but is still competitive with state-of-the-art SVS vocoders. 
Second, GOLF requires less than 40\% of memory to train and runs ten times faster than its alternatives for inference.
Moreover, we indirectly show that GOLF could model the phase components of the human voice by aligning the synthesised waveforms to the ground truth and calculating the differences.
This characteristic has excellent potential for analysing singing voice in a differentiable manner.
Decomposing the human voice into the glottal source and vocal tract could also enable us to adjust the singing style in different ways, such as altering the amount of vocal effort with varying shapes of the glottal pulse. 

\section{Background}
We first introduce the relevant notation.
$\x_i$ denotes the $i^{th}$ column vector and $x_{i,j}$ denotes the entry at the $i^{th}$ row and the $j^{th}$ column of the matrix $\mathbf{X}$.
Concatenating two matrices along the column dimension is denoted by $[; ]$.
$x_i$ denotes the $i^{th}$ entry of the vector $\x$ or a time sequence $x$ indexed by $i$.
$X(z)$ denotes the response of $x_n$ in the z-domain.
Unless stated otherwise, we use $n$ as the time index and $k$ as the frame index.
Angular frequencies and periods are normalised to the interval $[0, 1]$.
We use one-based indexing for elements with finite dimensions.

\subsection{Glottal Source-Filter Model}\label{sec:glottal_source}
In the source-filter model, we have the following simplified voice production model:
\begin{equation}
\label{eq:glottal_sf}
S(z) = \left (G(z) + N(z)\right )H(z)L(z),
\end{equation}
where $G(z)$ represents the periodic vibration from the vocal folds, $N(z)$ represents random components of the glottal source, $H(z)$ represents the vocal-tract filter, and $L(z)$ represents the radiation at the lips~\cite{degottex_glottal_nodate}.
Since this formulation is linear, the radiation filter $L(z)$ and the glottal pulse $G(z)$ can be merged into a single source $G'(z)$ called \emph{the radiated glottal pulse}.
If we assume $L(z)$ is a first-order differentiator  $1 - z^{-1}$~\cite{lu_glottal_nodate}, then $G'(z)$ is the derivative of the glottal pulse, which can be described by the LF model~\cite{fant_four-parameter_nodate}, a four-parameter model of glottal flow.
$H(z)$ is usually a Linear Predictive Coding (LPC) filter.

\subsection{Linear Predictive Coding}\label{sec:lpc}
LPC assumes that the current speech sample $s_n$ can be predicted from a finite number of previous $M$ samples $s_{n-1}$ to $s_{n-M}$ by a linear combination with residual errors $e_n$:
\begin{equation}
\label{eq:iir} 
s_n = e_n - \sum_{i = 1}^M a_i s_{n-i},
\end{equation}
where $a_i$ are the linear prediction coefficients.
This is the same as filtering the residuals, equivalent to the glottal source in our case, with an $M^{th}$-order all-pole filter, a filter that has an infinite impulse response (IIR).
We can use the LPC filter to represent the response of the vocal tract if the vocal tract is approximated by a series of cylindrical tubes with varying diameters~\cite {markel_linear_1976}, providing a physical interpretation.

Using LPC for neural audio synthesis is not new~\cite{valin_lpcnet_2019, oh_excitglow_2020, subramani_end--end_2022}, and works have been conducted to incorporate IIR filters and train them jointly with deep learning models~\cite{bhattacharya_optimization_nodate, nercessian_neural_nodate, kuznetsov_differentiable_nodate, colonel_direct_2021, nercessian_lightweight_2021, kim_joint_2022, steinmetz_style_2022}.
The difficulty of training IIR in deep learning framework (e.g.\ PyTorch) using \eqnref{eq:iir} is that its computation is recursive, i.e.\ the output at each step depends on the previous results, and to make the calculation differentiable, separated tensors are allocated in each step.
This generates a significant number of memory allocations and overheads for creating tensors, thus leading to performance issues, especially for long sequences.
One way to mitigate this is to allocate shared continuous memory before computation.
However, in-place modification is not differentiable in these frameworks.
Some studies sidestep the recursion by approximating IIR in the frequency domain using Discrete Fourier Transform (DFT)~\cite{oh_excitglow_2020, nercessian_neural_nodate, colonel_direct_2021, nercessian_lightweight_2021, kim_joint_2022, steinmetz_style_2022}, but the accuracy of this approximation depends on the DFT resolution.
Moreover, the IIRs used in practice are usually low-order; in this case, it is faster to compute them directly, especially on long sequences.

\section{Proposed Model}
Usually, $N(z)$ in \eqnref{eq:glottal_sf} is treated as amplitude-modulated Gaussian noise~\cite{lu_glottal_nodate, degottex_glottal_nodate}.
Our early experiments found this formulation to be challenging to optimise.
As an alternative, we move the noise components $N(z)$ outside the glottal source and filter it with time-varying filter $C(z)$, resulting in
\begin{equation}
\label{eq:glottal_hpn}
S(z) = G'(z)H(z) + N(z)C(z).
\end{equation}
This resembles the classic \emph{Harmonic-plus-Noise} model~\cite{serra_spectral_1990} and was used in previous DDSP-based SVS~\cite{alonso_latent_2021, wu1_ddsp-based_2022}.
Alonso et al.~\cite{alonso_latent_2021} modelled $G'(z)H(z)$ jointly using additive harmonic oscillators and time-varying Finite Impulse Responses (FIRs) as $C(z)$; Wu et al.~\cite{wu1_ddsp-based_2022} introduced a sawtooth oscillator as $G'(z)$ and zero-phase time-varying FIRs as $H(z)$.
In this work, we use a glottal flow model to synthesise harmonic sources and time-varying IIRs as filters.

 \subsection{Glottal Flow Wavetables}
\begin{figure}[h]
\centering
\includegraphics[width=\columnwidth]{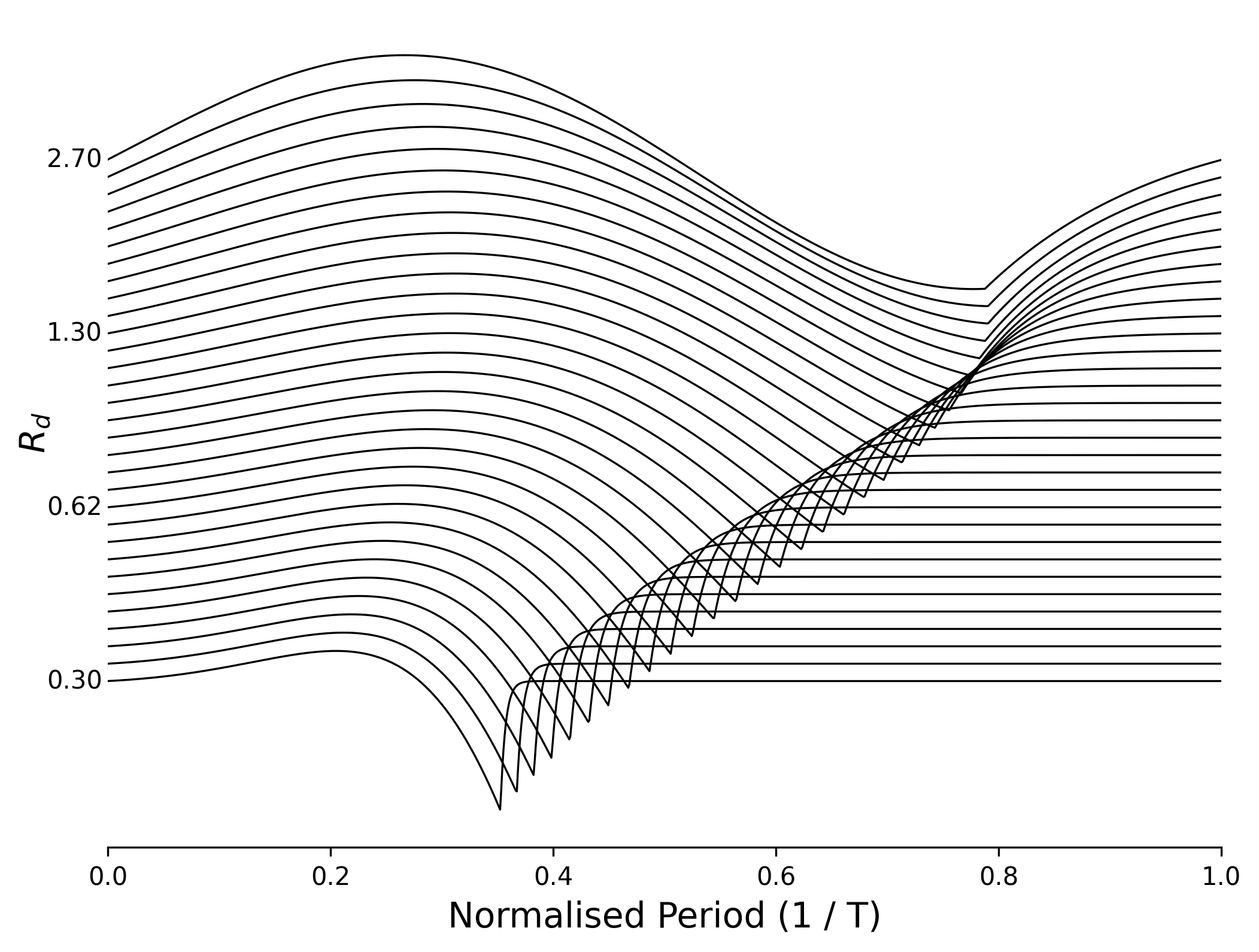}
\caption{An example of the wavetables we used, corresponding to matrix $\mathbf{D}$ with $K = 31$.}
\label{fig:glottal}
\end{figure}

We adopted the transformed-LF model~\cite{fant_lf-model_nodate} for generating glottal pulses.
This model re-parameterises the LF model~\cite{fant_four-parameter_nodate} using just one parameter $R_d$, which has been found to correspond to the perceived vocal effort well and covers a wide range of different glottal flow shapes.
We sampled $K$ values of $\log(R_d)$ with equal spacing inside $[\log(0.3), \log(2.7)]$ according to the value range suggested by \cite{degottex_glottal_nodate}.
We calculate the flow derivative function $g'(t; R_d)$ in continuous time $t$ for each sampled $R_d$ and then sampled $L$ points in one period to get its discrete version.
The details for calculating $g'(t; R_d)$ were given in \cite{gobl_reshaping_2017}.
By stacking these sampled glottal flows, we built wavetables $\mathbf{D} \in \RR ^{K \times L}$, with each row containing one period of a sampled glottal pulse (see Fig.~\ref{fig:glottal}).
The rows are sorted based on $R_d$.

The model generates glottal pulses $g'_n$ by linearly interpolating the two $\mathbf{D}$ axes.
The encoder network first predicts instantaneous frequency $f_n \in [0, 0.5]$ and the fractional index $\tau_n \in [0, 1]$ for $R_d$.
We then use the instantaneous phase $\phi_n = \sum_{i=1}^n f_i$ to interpolate the waveform as:
\begin{multline}
\label{eq:table_indexing}
g'_n = (1 - p) \left( (1 - q) \hat{d}_{\floor{k}, \floor{l}}
+ q \hat{d}_{\floor{k}, \ceil{l}}\right) \\
+ p \left( (1 - q) \hat{d}_{\ceil{k}, \floor{l}}
+ q \hat{d}_{\ceil{k}, \ceil{l}}\right),
\end{multline}
where $l = (\phi_n \mod 1)L + 1, k = \tau_n (K - 1) + 1, p = k - \floor{k}, q = l - \floor{l}$, and $\hat{\mathbf{D}} = [\mathbf{D}; \mathbf{d}_1] \in \RR^{K \times (L + 1)}$.
The wavetables $\mathbf{D}$ are fixed in our case, contrary to~\cite{shan_differentiable_2021}, and we only pick one wavetable at a time, not a weighted sum.

\subsection{Frame-Wise LPC Synthesis}
Time-varying LPC synthesis is usually done by linearly interpolating the LPC coefficients to the audio resolution and filtering sample by sample.
This is not parallelisable and slows down the training process.
As an alternative, we approximate LPC synthesis by treating each frame independently and using overlap-add:
\begin{equation}
\label{eq:lpc_ola}
s_n = \sum_k LPC(g'_n \gamma_n u_{n-kT}; \a_k)w_{n-kT},
\end{equation}
where $LPC(e_n; \mathbf{a})$ represents \eqnref{eq:iir}, $\a_k \in \RR^M$ are the filter coefficients at the $k^{th}$ frame, $u_n$ and $w_n$ are the windowing functions, $\gamma_n \in \RR^+$ is the gain, and $T$ is the hop size.
$u_n$ is fixed to the square window.
In this way, the computation can be parallelised. 
We found that the voice quality differences between overlap-add LPC and sample-by-sample LPC are barely noticeable if we use a sufficiently small hop size. 
We empirically found that a 200 Hz frame rate is sufficient.

\subsection{LPC Coefficients Parameterisation}
For the LPC filter to be stable, all of its poles must lie inside the unit circle on the complex plane.
Stability can be guaranteed using robust representations, such as reflection coefficients~\cite{subramani_end--end_2022}.
The representation we chose in this work is cascaded 2nd-order IIR filters, and we solve the stability issue by ensuring all the 2nd-order filters are stable.
We use the \emph{coefficient representation} from~\cite{nercessian_lightweight_2021} to parameterise the $i^{th}$ IIR filter's coefficients $1 + \eta_{i, 1}z^{-1} + \eta_{i, 2}z^{-2}$ from the encoder's outputs and cascade them together to form an $M^{th}$-order LPC filter:
\begin{equation}
\begin{split}
&  (1 + \eta_{1, 1}z^{-1} + \eta_{1, 2}z^{-2})(1 + \eta_{2, 1}z^{-1} + \eta_{2, 2}z^{-2}) \\
& \dotsb (1 + \eta_{\frac{M}{2}, 1}z^{-1} + \eta_{\frac{M}{2}, 2}z^{-2}) \\
& = 1 + a_1z^{-1} + a_2z^{-2} + \dotsb + a_Mz^{-M} = A(z).
\end{split}
\end{equation}

\subsection{Unvoiced Gating}\label{ssec:gating}

The instantaneous frequency $f_n$ predicted by the encoder is always non-zero and keeps the oscillator working.
Without constraint, the model would utilise these harmonics in the unvoiced region creating buzzing artefacts~\cite{wu1_ddsp-based_2022}.
We propose to mitigate this problem by jointly training the model to predict the voiced/unvoiced probabilities as $v_n \in [0, 1]$ and feeding the gated frequency $\hat{f}_n = v_n f_n$ to the oscillator instead.

\section{Optimisation}\label{sec:optimisation}
Training deep learning models is usually accomplished by backpropagating the gradients evaluated at a chosen loss function $\LL$ throughout the whole computational graph back to the parameters.
Partially inspired by Bhattacharya et al.\cite{bhattacharya_optimization_nodate}, we derived the closed form of \emph{backpropagation through time} to utilise efficient IIR implementation to solve the problems we mentioned in Section \ref{sec:lpc} while keeping the filter differentiable.
Here, $\mathbf{e} \in \RR^N$ is the input, $\a \in \RR^M$ is the filter coefficients, and $\mathbf{s} \in \RR^N$ is the output.
Assuming we know $\frac{\partial \LL}{\partial \mathbf{s}}$, we can get the derivatives $\frac{\partial \LL}{\partial \mathbf{e}}$ and $\frac{\partial \LL}{\partial \a}$, using chain rules $\frac{\partial \LL}{\partial \mathbf{s}}\frac{\partial \mathbf{s}}{\partial \mathbf{e}}$ and $\frac{\partial \LL}{\partial \mathbf{s}}\frac{\partial \mathbf{s}}{\partial \a}$.

\subsection{Backpropagation Through the Coefficients}\label{ssec:a_derivative}
Taking the derivatives of \eqnref{eq:iir} with respect to $a_i$ we get:
\begin{equation}
\label{eq:dsda}
\frac{\partial s_n}{\partial a_i} = -s_{n-i} - \sum_{k = 1}^M a_k \frac{\partial s_{n-k}}{\partial a_i},
\end{equation}
which equals $LPC(-s_{n-i}; \mathbf{a})$.
$s_n|_{n\leq0}$ does not depend on $a_i$ so the initial conditions $\frac{\partial s_n}{\partial a_i}|_{n\leq0}$ are zeros.
We can get $\frac{\partial s_n}{\partial \a}$ with one pass of filtering because $\frac{\partial s_n}{\partial a_j}$ is $\frac{\partial s_n}{\partial a_i}$ shifted by an offset $j - i$.
Lastly, we calculate $\frac{\partial \LL}{\partial a_i}$ as $\sum_{n=1}^N\frac{\partial \LL}{\partial s_n}\frac{\partial s_n}{\partial a_i}$.

\subsection{Backpropagation Through the Input}\label{ssec:x_derivative}
To get the derivatives for input $e_n$, we first re-write \eqnref{eq:iir} as the following convolutional form:
\begin{equation}
\label{eq:iir_conv}
s_n = \sum_{m = 1}^n e_m h_{n-m}, 
\end{equation}
where $h_{n} = \mathcal{Z}^{-1}\{H(z)\}, H(z) = \frac{1}{A(z)}$.
From \eqnref{eq:iir_conv} we see that $\frac{\partial s_n}{\partial e_m} = h_{n-m}$.
The derivative of loss $\LL$ with respect to $e_m$ depends on all future samples $s_n$, which is:
\begin{equation}
\label{eq:dLdem}
\begin{split}
\frac{\partial \LL}{\partial e_m} 
 = \sum_{n = m}^{N} \frac{\partial \LL}{\partial s_n}\frac{\partial s_n}{\partial e_m}
 = \sum_{n = m}^{N} \frac{\partial \LL}{\partial s_n} h_{n-m}.
\end{split}
\end{equation}
By swapping the variables $n,m$ and considering the equivalence of \eqnref{eq:iir} and \eqnref{eq:iir_conv}, \eqnref{eq:dLdem} can be simplified to 
\begin{equation}
\label{eq:dLde}
\begin{split}
\frac{\partial \LL}{\partial e_n} 
& = \sum_{m = n}^{N} \frac{\partial \LL}{\partial s_m} h_{m-n} \\ 
& = \frac{\partial \LL}{\partial s_n}  - \sum_{i = 1}^M a_i \frac{\partial \LL}{\partial e_{n+i}}.
\end{split}
\end{equation}
\eqnref{eq:dLde} shows that we can get the derivatives $\frac{\partial \LL}{\partial e_n}$ by just filtering $\frac{\partial \LL}{\partial s_n}$ with the same filter, but running in backwards.
The initial conditions $\frac{\partial \LL}{\partial e_n}|_{n > N}$ are naturally zeros.

In conclusion, backpropagation through an IIR filter consists of two passes of the same filter and one matrix multiplication\footnote{The computation of the IIR does not need to fulfil the implementation requirements set by the automatic differentiation framework, thus can be highly optimised.}. 
We implemented the IIR in C++ and CUDA with multi-threading to filter multiple sequences simultaneously\footnote{Although the single-core performance of a GPU is usually inferior to a CPU, and we can only use at most one thread for each IIR, the GPU has a much higher number of cores, which is beneficial for training on a large number of sequences at once.}.
The differentiable IIR is done by registering the above backward computation in PyTorch, and we submitted the implementation to TorchAudio~\cite{torchaudio} as part of the \texttt{torchaudio.functional.lfilter}.

\section{Experimental Setup}\label{sec:exp}
\subsection{Dataset}
We test GOLF as a neural vocoder on the MPop600 dataset~\cite{chu_mpop600_2020}, a high-quality Mandarin singing voice dataset featuring nearly 600 singing recordings with aligned lyrics sung by four singers.
We used the audio recordings from the \texttt{f1} (female) and \texttt{m1} (male) singers.
For each singer, we selected the first three recordings as the test set, the following 27 recordings as the validation set, and used the rest as training data (around three hours in total).
All the recordings were downsampled to 24 kHz.
The vocoder feature we choose is the log mel-spectrogram.
We computed the feature with a window size of 1024 and 80 mel-frequency bins and set the hop size $T$ to 120.
We normalised the feature to between zero and one and sliced the training data into two seconds excerpts with 1.5 seconds overlap.

\subsection{Model Details}
\begin{figure}[h]
\centering
\includegraphics[width=0.8\columnwidth]{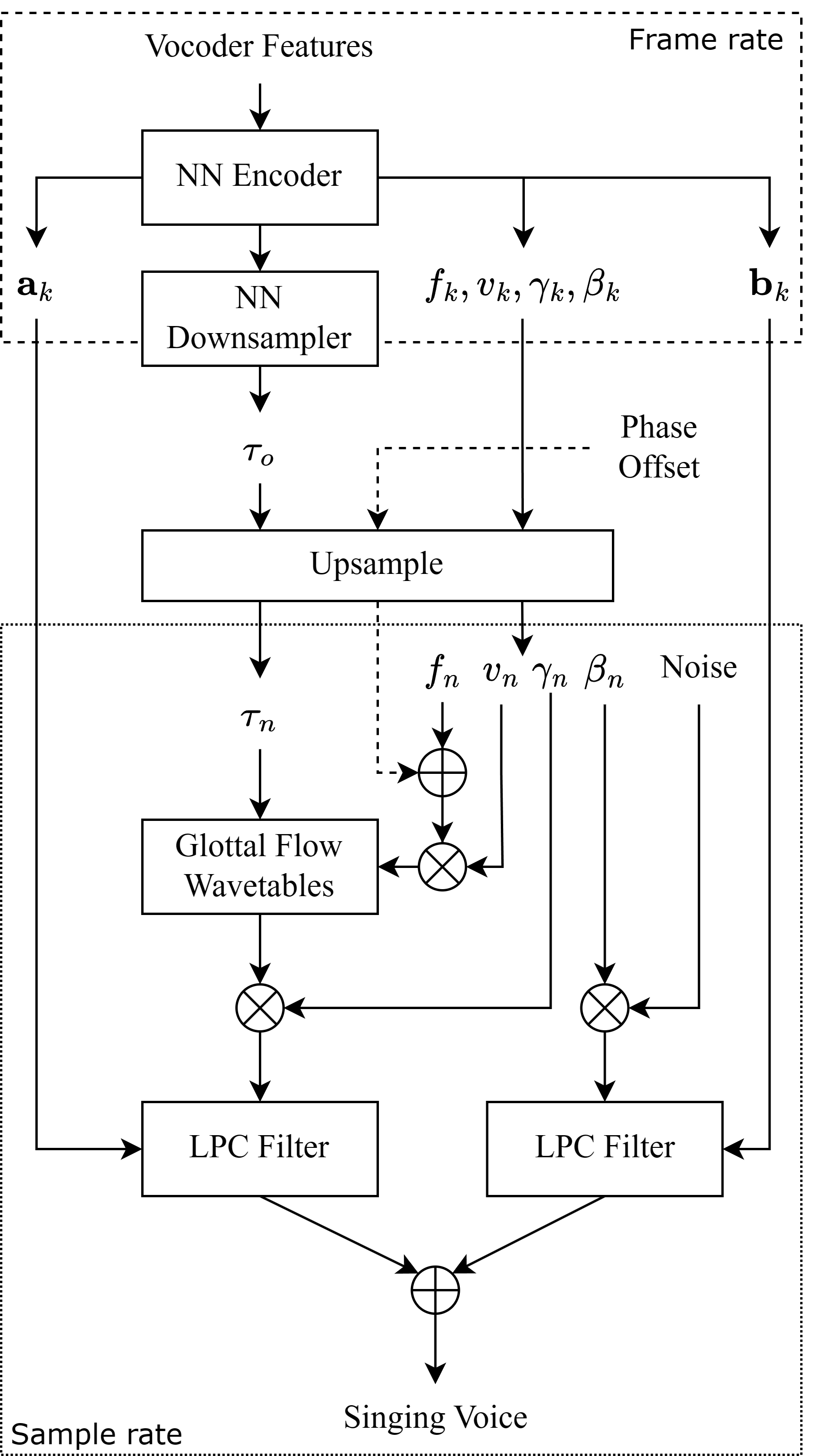}
\caption{Overview of the GOLF synthesis process. \emph{phase offset} is only introduced at test time, where the details are given in Section \ref{ssec:obj_eval}.}
\label{fig:diagram}
\end{figure}

We adopted the encoder from SawSing but replaced the transformer layers with three layers of Bi-LSTM for favourable implementation, resulting in around 0.7M parameters in total.
A final linear layer predicts the synthesis parameters $\{f_k, v_k, \gamma_k, \beta_k, \a_k, \mathbf{b}_k\}$.
The first four parameters are linearly upsampled to $\{f_n, v_n, \gamma_n, \beta_n\}$.
$\beta_n$ and $\mathbf{b}_k$ are the Gaussian noise's gain and filter coefficients.
We added an average pooling layer with a size of 10 and two convolution layers after the encoder to predict the $R_d$ fractional index $\tau_o$ at a lower rate and then linearly upsampled to $\tau_n$.
This step avoids possible modulation effects caused by switching the wavetables too quickly.
A system diagram of GOLF is shown in Fig.~\ref{fig:diagram}.
We set $K=100, L=2048$, Hanning window for $w_n$, and $M=22$ for both LPC filters.
We used the same hop size $T$ and a window size of 480 for frame-wise LPC.
We normalised all wavetables to have equal energy and aligned them with the negative peak.

\vspace{-0.2cm}

We compare GOLF with three DDSP-based baselines using the same NN encoder to predict their synthesis parameters.
The first two are the original DDSP~\cite{engel_ddsp_2020} and SawSing~\cite{wu1_ddsp-based_2022}.
We set their noise filter length to 80 and harmonic filter length to 256 for SawSing.
The third model is PUlse-train LPC Filter (PULF), which is similar to GOLF but replaces the glottal flow wavetables with a band-limited pulse train~\cite{nercessian_differentiable_2022} using additive synthesis, while the LPC order for the harmonic source is increased to 26 to accommodate the glottal pulse response.
The number of oscillating sinusoids was set to over 150 for all the baselines.
We did not compare GOLF with Nercessian et al.~\cite{nercessian_differentiable_2022} and Yoshimura et al.~\cite{yoshimura_embedding_2022} because these architectures are based closely on the source-filter model, and use additional post-nets to enhance the voice, which makes it harder to compare directly with GOLF.

\subsection{Training Configurations}
We trained separate models for each singer, resulting in 8 models.
The loss function is the summation of the multi-resolution STFT loss (MSSTFT) and f0 loss from SawSing with FFT sizes set to $\{512, 1024, 2048\}$, plus a binary cross entropy loss on voiced/unvoiced prediction.
We stopped the gradients from the harmonic source to the f0s and voiced decisions to stabilise the training.
We used Adam~\cite{kingma_adam_2017} for running all optimisations.
For DDSP and SawSing, the batch size and learning rate were set to 32 and 0.0005; for GOLF and PULF, the numbers were 64 and 0.0001.
We used the ground truth f0s (extracted by WORLD~\cite{morise_world_2016}) for the harmonic oscillator of PULF during training due to stability issues.
We trained all the models for 800k steps to reach sufficient convergence and picked the checkpoint with the lowest validation loss as the final model\footnote{The trained checkpoints, source codes, and audio samples are available at \url{https://github.com/iamycy/golf}.}.

\section{Evaluations}

\subsection{Objective Evaluation}
\label{ssec:obj_eval}
The objective metrics we choose are the MSSTFT, the mean absolute error (MAE) in f0, and the Fr{\'e}chet audio distance (FAD)~\cite{kilgour_frechet_2019} on the predicted singing of the test set.
Table~\ref{tab:comparison} shows that DDSP has the lowest MSSTFT and f0 errors, while SawSing reaches the lowest FAD.
GOLF and PULF show comparable results in f0 errors to other baselines.
We report the memory usage when training these models and their real-time factor (RTF), both on GPU and CPU, in Table~\ref{tab:rtf}. 
The amount of memory required to train GOLF is around 35\% of others, and it runs extremely fast, especially on the CPU.

\begin{table}[h]
\centering
\resizebox{\columnwidth}{!}{%
\begin{tabular}{@{}ccccc@{}}
\toprule
Singers                      & Models  & MSSTFT          & MAE-f0 (cent)           & FAD                       \\ \midrule
\multirow{4}{*}{\texttt{f1}} & DDSP    &\textbf{3.09}    & \textbf{74.47}$\pm$1.19 & 0.50$\pm$0.02             \\
                             & SawSing & 3.12            &  78.91$\pm$1.18         & \textbf{0.38}$\pm$0.02    \\ \cmidrule(l){2-5} 
                             & GOLF    & 3.21            &  77.06$\pm$0.88         & 0.62$\pm$0.02             \\
                             & PULF    & 3.27            &  76.90$\pm$1.11         & 0.75$\pm$0.04             \\ \midrule
\multirow{4}{*}{\texttt{m1}} & DDSP    & \textbf{3.12}   & \textbf{52.95}$\pm$1.03 & 0.57$\pm$0.02             \\
                             & SawSing & 3.13            &  56.46$\pm$1.04         & \textbf{0.48}$\pm$0.02    \\ \cmidrule(l){2-5} 
                             & GOLF    & 3.26            &  54.09$\pm$0.30         & 0.67$\pm$0.01             \\
                             & PULF    & 3.35            &  54.60$\pm$0.73         & 1.11$\pm$0.04             \\  \bottomrule
\end{tabular}%
}
\caption{Evaluation results on the test set. We omit the standard deviation if it is smaller than 0.01.}
\label{tab:comparison}
\end{table}

As an additional metric we use the L2 loss between the predicted and the ground truth waveform.
The intuition behind this is that GOLF and PULF are the only two models introducing non-linear phase response because of IIR filtering.
The filters in DDSP and SawSing are all zero-phase, and the initial phases of the sinusoidal oscillators are fixed to zeros.
We emphasise that this test is not targeting human perception but the phase reconstruction ability of the models. Humans cannot perceive the absolute frequency phase, but accurate reconstruction could be important in sound matching and mixing use cases.
We evaluate the loss on one of the test samples from \texttt{m1} we used in the subjective evaluation.
We created a new parameter called \emph{phase offset} sampled at 20 Hz.
We linearly upsampled \emph{phase offset} and added it to the instantaneous phase $\phi_n$, introducing a slowly varying phase shift.
We optimised this parameter by minimising the predicted waveform's L2 loss to the ground truth using Adam with a learning rate of 0.001 and 1000 steps.
We wrapped the differences between the points of \emph{phase offset} during optimisation to [-0.5, 0.5].
We ran this optimisation five times for each model. 
Each time the \emph{phase offset} was initialised randomly.
We report the minimum and maximum final losses from these trials.
Table~\ref{tab:rtf} shows the lowest losses GOLF and PULF can reach are significantly smaller than the others, with GOLF having the smallest among all.

\begin{table}[h]
\centering
\resizebox{0.9\columnwidth}{!}{%
\begin{tabular}{@{}cccccc@{}}
\toprule
\multirow{2}{*}{Models} & \multirow{2}{*}{Memory} & \multicolumn{2}{c}{RTF}         & \multicolumn{2}{c}{Waveform L2} \\ \cmidrule(l){3-6} 
        &     & GPU   & CPU   & Min   & Max   \\ \midrule
DDSP    & 7.3 & 0.015 & 0.237 & 71.83 & 88.77 \\
SawSing & 7.3 & 0.015 & 0.240 & 75.72 & 93.16 \\
GOLF    & \textbf{2.6}            & \textbf{0.009} & \textbf{0.023} & \textbf{21.98} & \textbf{64.82} \\
PULF    & 7.5 & 0.015 & 0.248 & 44.08 & 70.59 \\ \bottomrule
\end{tabular}%
}
\caption{The required number of VRAM (GB) for training with a batch size of 32, real-time factor (RTF), and the minimum/maximum L2 loss on waveform using one of the test samples. 
The benchmark was conducted on an Ubuntu 20.04 LTS machine with an i5-4790k processor and an NVIDIA GeForce RTX 3070 GPU.}
\label{tab:rtf}
\end{table}

\begin{figure*}[t!]
\centering
\includegraphics[width=\textwidth]{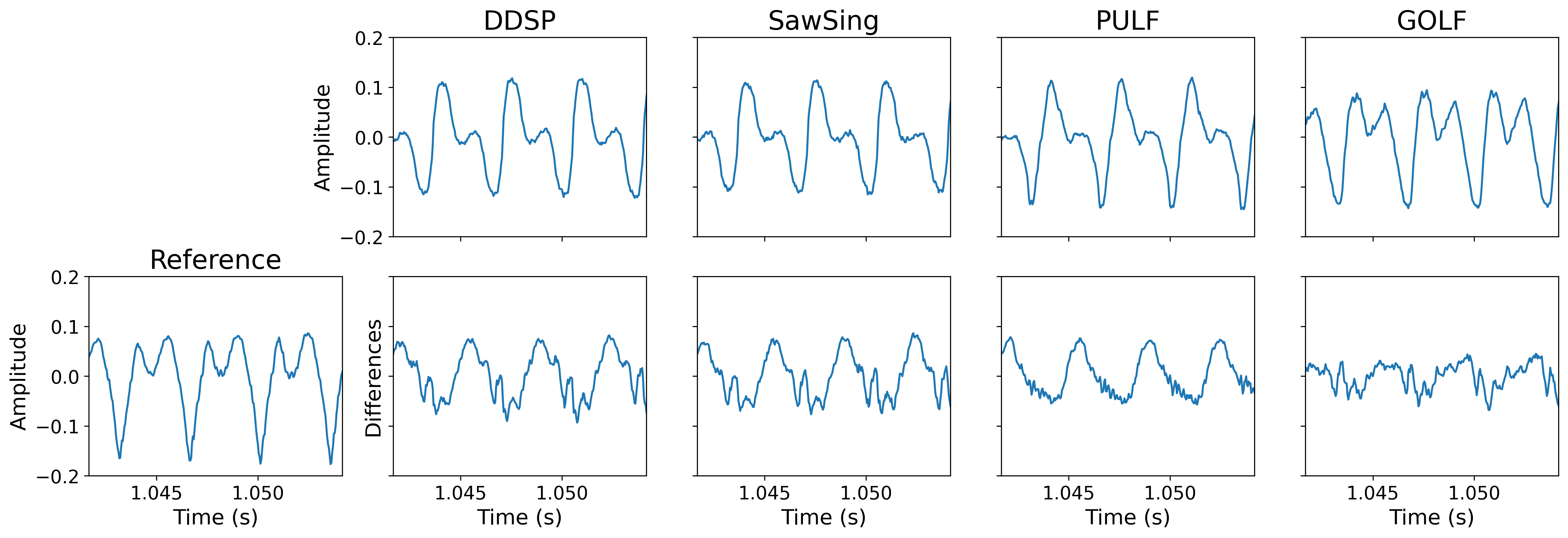}
\caption{The predicted waveforms of a short segment from one of the \texttt{m1} test samples. The differences were computed by subtracting the predicted signal from the reference.}
\label{fig:wav_comparison}
\end{figure*}

\subsection{Subjective Evaluation}
\begin{figure}[h]
\centering
\includegraphics[width=\columnwidth]{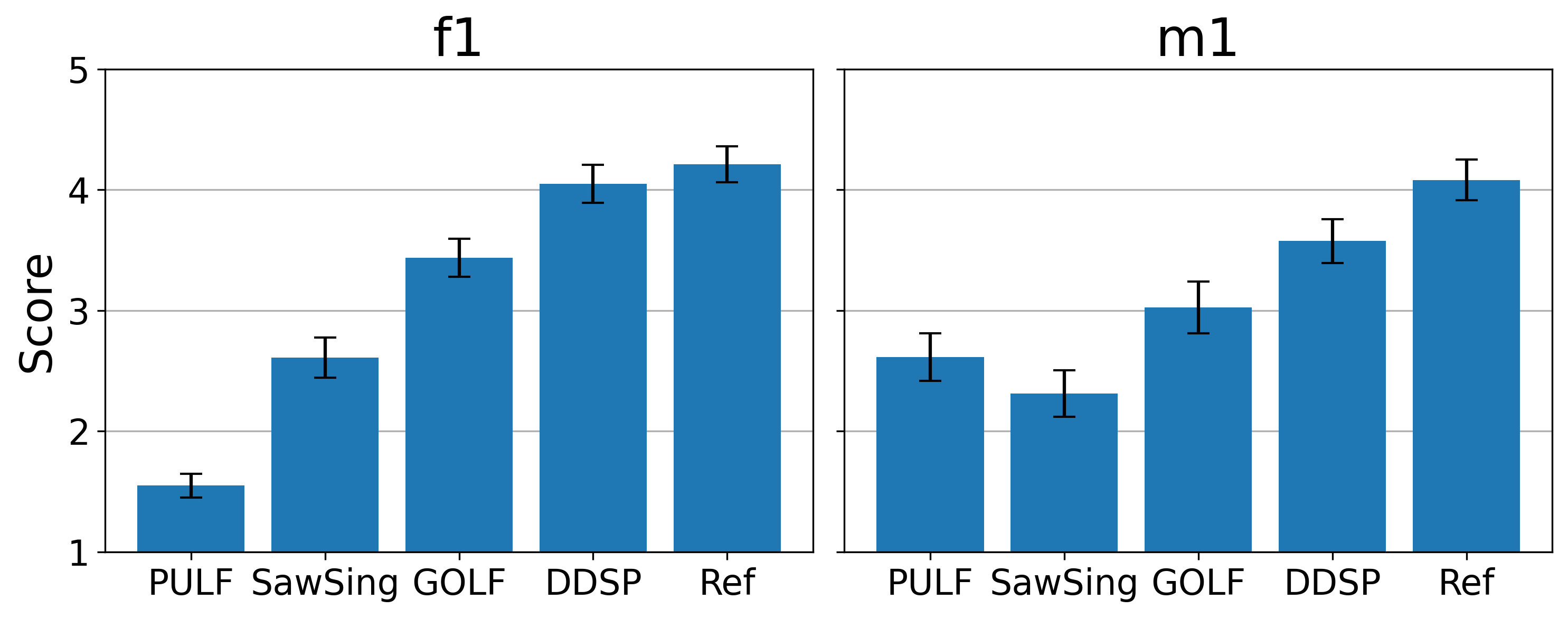}
\caption{The MOS results of the vocoders trained on different singers with 95\% confidence interval.}
\label{fig:mos}
\end{figure}

We conducted an online listening test using Go Listen~\cite{hines_go_2021}.
We picked one short clip from each test set recording, resulting in 6 clips with duration ranging from 6 to 11 seconds.
The test is thus divided into six examples, each consisting of one ground truth clip and four synthesised results from different models, and their loudness was normalised to –16dB LUFS.
The order of the examples and the stimulus were randomised for each subject.
Each subject was requested to rate the quality of these stimuli on a score from 0 to 100.
We collected responses from 33 anonymous participants.
We dropped one of the participants who did not indicate using headphones.
We normalise scores to fall between 1 to 5 and report the Mean Opinion Score (MOS) in Fig.~\ref{fig:mos}.
DDSP has the highest opinion scores overall, and a Wilcoxon signed-rank test shows that it is not statistically significantly different from the \texttt{f1} ground truth (\emph{p} = 0.168).
We applied the one-side Wilcoxon test on GOLF and PULF to compare them with SawSing, and the results show that GOLF significantly outperforms SawSing (\emph{p} < 0.0001), and PULF performs better than SawSing on \texttt{m1} (\emph{p} < 0.022).

\section{Discussions}\label{sec:discussions}
Given the evaluation results and the number of synthesis parameters in GOLF is roughly six times smaller than DDSP and SawSing, it is clear that GOLF's synthesis parameters are a more compact representation.
Comparing the differences between GOLF and PULF in Table~\ref{tab:rtf}, we can see that the performance gain is due to the use of wavetables.
Other baselines synthesise band-limited harmonic sources with many sinusoids oscillating simultaneously, thus increasing the computational cost.
PULF's MOS score is much worse on \texttt{f1}, with noticeable artefacts in the unvoiced region and random components of the voice.
After investigation, we found the noise gains $\beta_n$ predicted by PULF fluctuating at high speed, producing a modulation effect.
This behaviour is also found in PULF trained on \texttt{m1} and GOLF, even on the harmonic gains $\gamma_n$. 
Still, the amount of fluctuation is small and barely noticeable in the test samples.
Given the available results, we could only conclude that this effect relates to the type of harmonic source and the range of f0, i.e., female singers have higher f0.
This amplitude modulation effect cannot be observed in spectrograms and thus is not captured by the training loss we used. 
It could be an intrinsic drawback of using frame-wise LPC approximation, but more experiments and comparisons with sample-wise LPC are needed.
In addition, SawSing produced low scores for both singers because of the buzzing artefacts in the unvoiced region. 
Although unvoiced gating (Sec.~\ref{ssec:gating}) reduces this problem to a large degree, human ears are susceptible to this effect.
This could be an inherent problem in using a sawtooth as the harmonic source.

The L2 loss shown in Table~\ref{tab:rtf} demonstrates that GOLF matches phase-related characteristics more accurately than other models.
Fig.~\ref{fig:wav_comparison} shows GOLF produces the most similar waveform to the ground truth.
Other baselines' waveforms are similar because they use the same additive synthesiser.
It is possible to reduce their L2 loss by optimising the initial phases of the oscillators, but this cannot account for time-varying source shapes.
Low L2 loss is a positive effect of the deterministic phase responses embedded in GOLF.
This opens up many possibilities, such as decomposing and analysing the voice in a differentiable manner and training the vocoder using the time domain loss function. 
The latter could be a possible way to reduce the fluctuation problem discussed in the previous paragraph.
The waveform matching of GOLF can be improved further by using a more flexible glottal source model, adding FIR and all-pass filters to account for the voice's mixed-phase components and the recording environment's acoustic response.

Lastly, we note that cascaded IIR filters provide an orderless representation (i.e.\ the cascading order does not affect the outputs).
This results in the \emph{responsibility problem}~\cite{zhang_deep_2020, hayes_responsibility_2023} for the last layer of the encoder, which might be one of the reasons why GOLF and PULF are less stable to train than other baselines.
Developing architectures that can handle orderless representation or switch to other robust representations are possible ways to address this.

\section{Conclusions}\label{sec:conclusions}
We present a lightweight singing voice vocoder called GOLF, which uses wavetables with different glottal flows as entries to model the time-varying harmonic components and differentiable LPC filters for filtering both the harmonics and random elements.
We show that GOLF requires less memory to train and runs an order of magnitude faster on the CPU than other DDSP-based vocoders, but still attains competitive voice quality in subjective and objective evaluations.
Furthermore, we empirically show that the predicted waveforms from GOLF represent the voice's phase response more faithfully, which could allow us to use GOLF to decompose and analyse human voice.

\section{Acknowledgements}

The authors want to thank Moto Hira (\texttt{mthrok}), Christian Puhrsch (\texttt{cpuhrsch}), and Alban Desmaison (\texttt{albanD}) for reviewing our pull requests to TorchAudio.
We are incredibly grateful to Parmeet Singh Bhatia (\texttt{parmeet}) for implementing the first version of efficient IIR in the TorchAudio codebase as \texttt{lfilter.cpp}.
We thank Ben Hayes for giving feedback on the equations of backpropagation through an IIR.
The first author wants to exclusively thank Ikuyo Kita for giving positive, energetic support during the writing process.
The first author is a research student at the UKRI Centre for Doctoral Training in Artificial Intelligence and Music, supported jointly by UK Research and Innovation [grant number EP/S022694/1] and Queen Mary University of London.

% For bibtex users:
\bibliography{clear}

\end{document}